# Improving VERITAS Sensitivity by Fitting 2D Gaussian Image Parameters

Jodi Christiansen[a] for the VERITAS Collaboration

[a]*California Polytechnic State University, San Luis Obispo, CA 93407.*

**Abstract.** Our goal is to improve the acceptance and angular resolution of VERITAS by implementing a camera image-fitting algorithm. Elliptical image parameters are extracted from 2D Gaussian distribution fits using a $\chi^2$ minimization instead of the standard technique based on the principle moments of an island of pixels above threshold. We optimize the analysis cuts and then characterize the improvements using simulations. We find an improvement of 20% less observing time to reach 5-sigma for weak point sources.

## INTRODUCTION

VERITAS [1][2], located at the Fred Lawrence Whipple Observatory in southern Arizona, USA, is an array of four 12 meter diameter imaging atmospheric Cherenkov telescopes (IACTs). Showers caused by high-energy gamma rays that interact in the atmosphere are imaged in the camera as shown in Figure 1. These images are then used to reconstruct the incident gamma-ray direction and energy. Image reconstruction algorithms have a variety of issues to deal with. First, especially for energetic events, it is possible that only part of the shower is imaged in the camera. Second, pixels may be removed for a variety of hardware failure modes or because of contamination from bright stars that are imaged at the same location in the camera. In these cases we find that fitting a 2D Gaussian distribution to the data gives better directional accuracy. The blue ellipse in Figure 1 is a fitted 2D Gaussian distribution and the red ellipse is produced with the standard principle component technique.

## ESTIMATING HILLAS PARAMETERS

Hillas parameters [3] define the average location of the image centroid ($x_c, y_c$), its length, width, orientation, and integrated photoelectrons (size). The standard technique of estimating the Hillas parameters calculates the moments of an island of pixels above some cleaning threshold, typically a few photoelectrons [4].

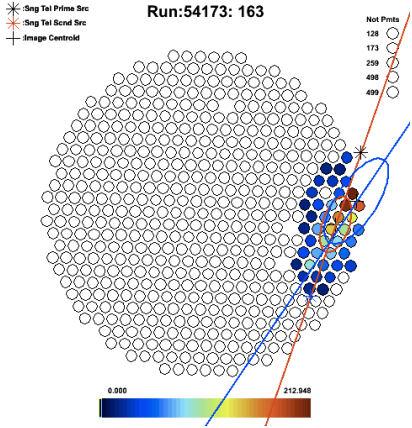

**FIGURE 1.** Camera display of an event showing pixels that are above threshold. The light from this air shower extends beyond the edge of the camera. The blue ellipse is a fitted 2D Gaussian distribution and the red ellipse is produced with the standard principle component technique.

The camera image-fitting method finds the optimal parameters of a 2D Gaussian distribution.

$$G(x_i, y_i) = A \exp\left(-\frac{1}{2}\left(\frac{(x_i - x_c)\cos\theta - (y_i - y_c)\sin\theta}{\sigma_{length}}\right)^2 - \frac{1}{2}\left(\frac{(x_i - x_c)\sin\theta + (y_i - y_c)\cos\theta}{\sigma_{width}}\right)^2\right) \quad (1)$$

where each pixel has a location $(x_i, y_i)$, and the parameters are the amplitude $A$, the centroid $(x_c, y_c)$, the length and width defined by $\sigma_{length}$ and $\sigma_{width}$, and the $\theta$ orientation of the image in the camera defined with respect to the radial axis.

A minimization between the 2D Gaussian and the data relies on the $\chi^2$ estimator.

$$\chi^2 = \frac{1}{N_{DOF}}\left(\sum_{all\ pixels} \frac{(S_i - G(x_i, y_i))^2}{\sigma_i^2}\right) \quad (2)$$

where the sum is over each pixel in the camera with signal, $S_i$, and the statistical uncertainty, $\sigma_i^2$, is estimated as the squared sum of the phototube pedestal variance and the statistical counting uncertainty from the expected signal, $G(x_i, y_i)$. It's important to seed the fitting routine with a reasonable guess so that the minimum gradient method will steadily converge to the optimal parameter estimation. We use the standard Hillas parameters to seed the algorithm. We find that the fitted parameters have better directional resolution than the standard parameters. The only drawback is that the fitting algorithm takes an additional 20 msec per event to process. This is about 15 times longer than the standard moment analysis.

## GAMMA-RAY RECONSTRUCTION

The impact point of the gamma ray shower axis projected to the ground, the zenith angle, and azimuth are determined from the centroid and orientation of the images in the camera parameterized by the Hillas parameters. Stereoscopic reconstruction requires detections in two or more telescopes. The gamma-ray energy reconstruction also depends on the integrated signal defined by the Hillas parameters as well as the distance from the shower to the telescope.

Cuts are used to differentiate between cosmic-ray backgrounds and VHE gamma-ray signals. Table 1 lists the image cuts for both algorithms optimized for a dim source with a spectral index similar to the Crab Nebula. To compare the two algorithms for determining the Hillas parameters, we had to determine the size cut that resulted in the same gamma-ray energy threshold for the two algorithms. The size in the standard analysis is simply the sum over all the signals, $S_i$, in the pixels associated with an island of 5 or more pixels that are above a fixed threshold. This differs from the integral over the fitted 2D Gaussian kernel.

Another difference is the fiducial cut. For the standard analysis, a cut on the distance (dist) of the image centroid from the center of the camera is used to ensure that the image is not clipped at the edge of the camera. For the 2D Gaussian analysis we instead require that the ratio of the standard size to the extrapolated size of the 2D Gaussian (sizeFrac) is greater than 25%. This ensures that enough of the shower image is contained within the camera to determine the Hillas parameters. The cuts on the reconstructed air showers turned out to be the same for both optimized analyses. These cuts include mean scaled width (MSW<1.15°), means scaled length (MSL<1.3°), the height of shower maximum (shwrMax>7 km), and $\Delta\theta<0.1°$.

TABLE 1. Image cuts used for analysis

| Standard Analysis | 2D Gaussian Fit |
|---|---|
| 5 pixels with signal | Seed with standard analysis Hillas parameters (5 pixels with signal) |
| size>400 | fitSize>420 |
| dist<1.43° | sizeFrac>0.25 |

## IMPACT ON ANALYSIS

Simulations are used to demonstrate the benefit of the new algorithm. Figure 2a shows that 2D Gaussian fit produces larger effective areas. This is due to the fact that the fits can extrapolate beyond the edge of the camera whereas the standard analysis is biased by the edge of the camera and requires stiffer fiducial cuts. Figure 2b shows that the angular resolution is also better especially at high energy. The improved fitted image axis in the camera leads to a better angular resolution for the reconstructed gamma ray. These two effects result in a better sensitivity for dim sources of about 20% as shown in Figure 3.

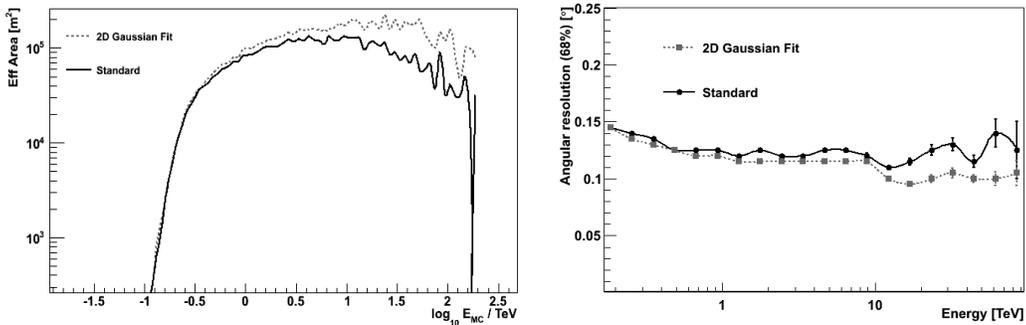

**FIGURE 2.** (a) The effective area of the array is increased by improving the fiducial cut in the camera. (b) The 2D Gaussian fit method improves the angular resolution of the array for high-energy gamma-rays. This simulation is for a zenith angle of 20°.

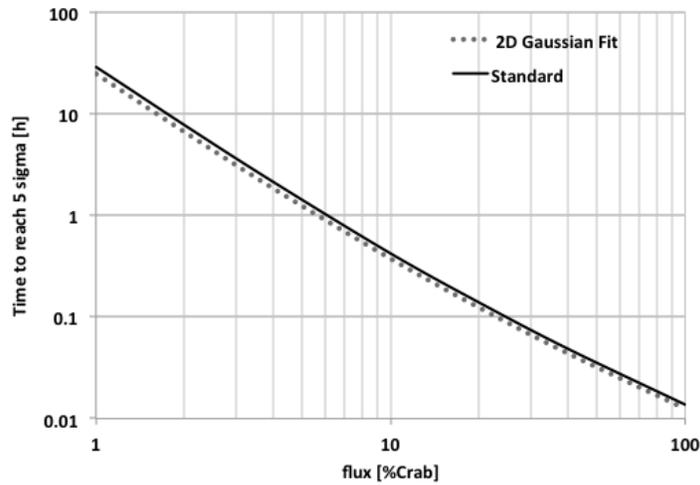

**FIGURE 3.** Observing time in hours required to reach 5$\sigma$ for a source flux in Crab Nebula units.

## ACKNOWLEDGMENTS

This research is supported by grants from the U.S. Department of Energy Office of Science, the U.S. National Science Foundation and the Smithsonian Institution, by NSERC in Canada, by Science Foundation Ireland (SFI 10/RFP/AST2748) and by STFC in the U.K. We acknowledge the excellent work of the technical support staff at the Fred Lawrence Whipple Observatory and at the collaborating institutions in the construction and operation of the instrument.